\documentclass[%
 reprint,
superscriptaddress,
 amsmath,amssymb,
 aps,
pra,
]{revtex4-2}

\usepackage{graphicx}
\usepackage{dcolumn}
\usepackage{bm}

\usepackage{physics}
\usepackage{comment}

\begin{document}

\preprint{APS/123-QED}

\title{Transition from laser Raman to Bernstein scattering in magnetized plasmas}

\author{Ryan Y. Lau}
\email{ryan.lau@colorado.edu}
\affiliation{Department of Physics, Center for Integrated Plasma Studies, University of Colorado, Boulder, Colorado 80309, USA}
\author{David J. Strozzi}
\affiliation{Lawrence Livermore National Laboratory, P.O. Box 808, Livermore California 94551-0808, USA}
\author{John D. Moody}
\affiliation{Lawrence Livermore National Laboratory, P.O. Box 808, Livermore California 94551-0808, USA}
\author{John R. Cary}
\affiliation{Department of Physics, Center for Integrated Plasma Studies, University of Colorado, Boulder, Colorado 80309, USA}
\affiliation{Silvaco Group, Inc., 5621 Arapahoe Avenue, Boulder, CO 80303}
\author{Yuan Shi}
\email{Yuan.Shi@colorado.edu}
\affiliation{Department of Physics, Center for Integrated Plasma Studies, University of Colorado, Boulder, Colorado 80309, USA}

\date{\today}

\begin{abstract}
When a laser backscatters from an external perpendicular magnetic field, stimulated Raman scattering transitions to stimulated Bernstein scattering, which we show using kinetic theory and particle-in-cell simulations. 
For low field strengths, the magnetic field acts as an electron detrapping mechanism, suppressing nonlinear effects of Raman scattering.  
At stronger fields, when the electron cyclotron frequency $\omega_{\text{ce}}$ becomes non-negligible compared to the plasma frequency $\omega_{\text{pe}}$, 
the Raman resonance splits into discrete resonances mediated by electron Bernstein waves (EBWs). 
Raman scattering ceases to exist between the EBW resonances, which are narrower and stronger than the Raman resonance they replace.  
The replacement is sensitive to angle: When the magnetic field is away from perpendicular by an angle $\alpha \gtrsim \omega_{\text{ce}}/\omega_{\text{pe}}$, the damping of EBW resonances blends them back into a broad single Raman peak. 
Our results demonstrate that strong magnetic fields, either spontaneously generated or externally imposed in high-energy-density experiments, fundamentally change laser-plasma interactions in a way that has no analog in unmagnetized plasmas, with implications for field diagnostics, plasma photonics, and inertial confinement fusion. 
\end{abstract}

\maketitle


Megagauss (MG) magnetic fields are common in high-energy-density plasmas. Even without a seed field, laser-produced plasmas often generate Biermann battery fields and Weibel filaments, which are subsequently transported and amplified due to internal plasma dynamics \cite{stamper1971spontaneous, weibel1959instability, huntington2015observation, Farmer17, Walsh_2021, appelbe2021magnetic, oneill2025burn, Lezhnin26}.
Moreover, externally imposed seed fields can improve the performance of laser-driven X-ray sources \cite{Li2019,Filippov2021} and inertial confinement fusion (ICF) \cite{Chang11, Hohenberger12, schmit2014understanding, moody2022increased} by modifying instabilities, improving particle confinement, and suppressing thermal conduction.
When seed magnetic fields undergo flux compression, tens of MG fields have been inferred in experiments \cite{Knauer10, Sio21, Bailly-Grandvaux2024}.

The roles of magnetic fields on long-pulse laser-plasma interactions (LPIs) are multifold \cite{shi2018laser} and are not well understood compared to unmagnetized LPIs \cite{michel2023introduction}. 
In addition to indirect effects due to changing plasma conditions \cite{Froula07,Montgomery15}, magnetic fields modify the Langmuir wave, which mediates two-plasmon decays \cite{Barr84,Li23TPD,montgomery2016two} and Raman scattering \cite{Nabil98,Paknezhad16,Jia17,Winjum18,Bolanos25,Lee26}, and also modifies the ion-acoustic wave, which mediates Brillouin scattering \cite{Karmendy79,Grebogi80,Brodin13,Shi19} and crossbeam energy transfer \cite{MoodyLinTheory, Shi26}. 
Moreover, magnetized plasmas support a zoo of waves, which can mediate additional resonances \cite{Shi17three, Shi17compression,Edwards19,Manzo22}.

In this letter, we show that lasers can backscatter from electron Bernstein waves (EBWs), which replace, rather than coexist with, the usual Raman scattering.
Scattering from EBWs has long been theorized \cite{Platzman68,Sharma80,Stenflo81,Liu1986parametric,Simon95} but in isolation from other processes. We demonstrate the transition from Raman to Bernstein scattering using PIC simulations.  
For lasers with vacuum wavelength \mbox{$\lambda_0=351$ nm} and under typical ICF conditions, the transition occurs when $B_0\gtrsim 1$ MG, or when $B_0\gtrsim 100$ T, and only for propagation angles within $\alpha\lesssim \omega_{\text{ce}}/\omega_{\text{pe}}$ from perpendicular to the magnetic field, where $\omega_{\text{ce}}$ is the electron cyclotron frequency and $\omega_{\text{pe}}$ is the plasma frequency.
As $\omega_{\text{ce}}$ increases, a single broad Raman resonance splits into a spectrum of narrower but stronger EBW resonances, which may complicate the use of broadband lasers to mitigate Raman scattering \cite{deeney2022advances,Kang_2025}. 
Moreover, the sensitivity of the spectrum to both the strength and the direction of the magnetic field may enable the use of laser scattering to measure MG fields in localized regions, which are difficult to infer using path-integrated diagnostics like proton \cite{Bott17,Schaeffer23} and electron \cite{Schumaker13,Zhang22} radiography. 

\textit{Methods.} Our Vlasov theory of magnetized LPIs is described in Ref.~\cite{MoodyLinTheory}. We calculate the gain spectrum of stimulated backscattering, where the pump (\mbox{$\lambda_0=351$ nm}) and the seed ($\lambda_1 = \lambda_0 + \Delta \lambda$) lasers propagate along the $x$ direction. 
To isolate magnetization effects, we fix the plasma density at $n_e = Zn_i = 10^{20}\,\text{cm}^{-3}$ and the temperature at $T_e=T_i=400$ eV, where the ion species is fully ionized carbon ($Z=6$). The distribution functions are initially Maxwellian. 
The external magnetic field is $\mathbf{B} = B_0\left(\vu{z}\sin\theta + \vu{x}\cos\theta\right)$, where $B_0$ is uniform. When $\theta=90^\circ$, the field is perpendicular to the laser propagation direction.

Our protocol of using particle-in-cell (PIC) simulations to extract the gain spectrum of laser scattering is described in Ref.~\cite{shi2023benchmarking, Shi26}. Stimulated scattering, rather than spontaneous scattering seeded by PIC noise, gives a larger signal to noise ratio. 
We use the 1D-3V version of the relativistic EPOCH code \cite{arber2015contemporary} with collisions turned off and plasma parameters identical to the theory. 
The length of the simulation domain is $L_x=500\lambda_0$.
Within two vacuum gaps of $10\lambda_0$ on either side of the domain, the plasma density is constant. 
We launch the pump ($I_{0} = 10^{15}\, \text{W/cm}^2$) and seed ($I_{1} = 10^{13}\, \text{W/cm}^2$) lasers from the right and left domain boundaries, respectively. 
From PIC data, we use Faraday filters to separate right- and left-moving waves, a Fourier filter to reduce PIC noise, and the Hilbert transform to obtain pump and seed envelopes, from which we calculate the seed gain.
Numerical resolutions are scanned to ensure convergence. 
The results shown below are for both lasers polarized along the $z$ direction, which corresponds to the O mode when $\theta=90^\circ$. The results for X modes are similar. 
When $\theta$ decreases away from $90^\circ$, the lasers launched in PIC simulations are predominantly O modes, but are mixed with an increasing amount of X modes.

\begin{figure} [t]
\begin{center}
\includegraphics[scale=1.0]{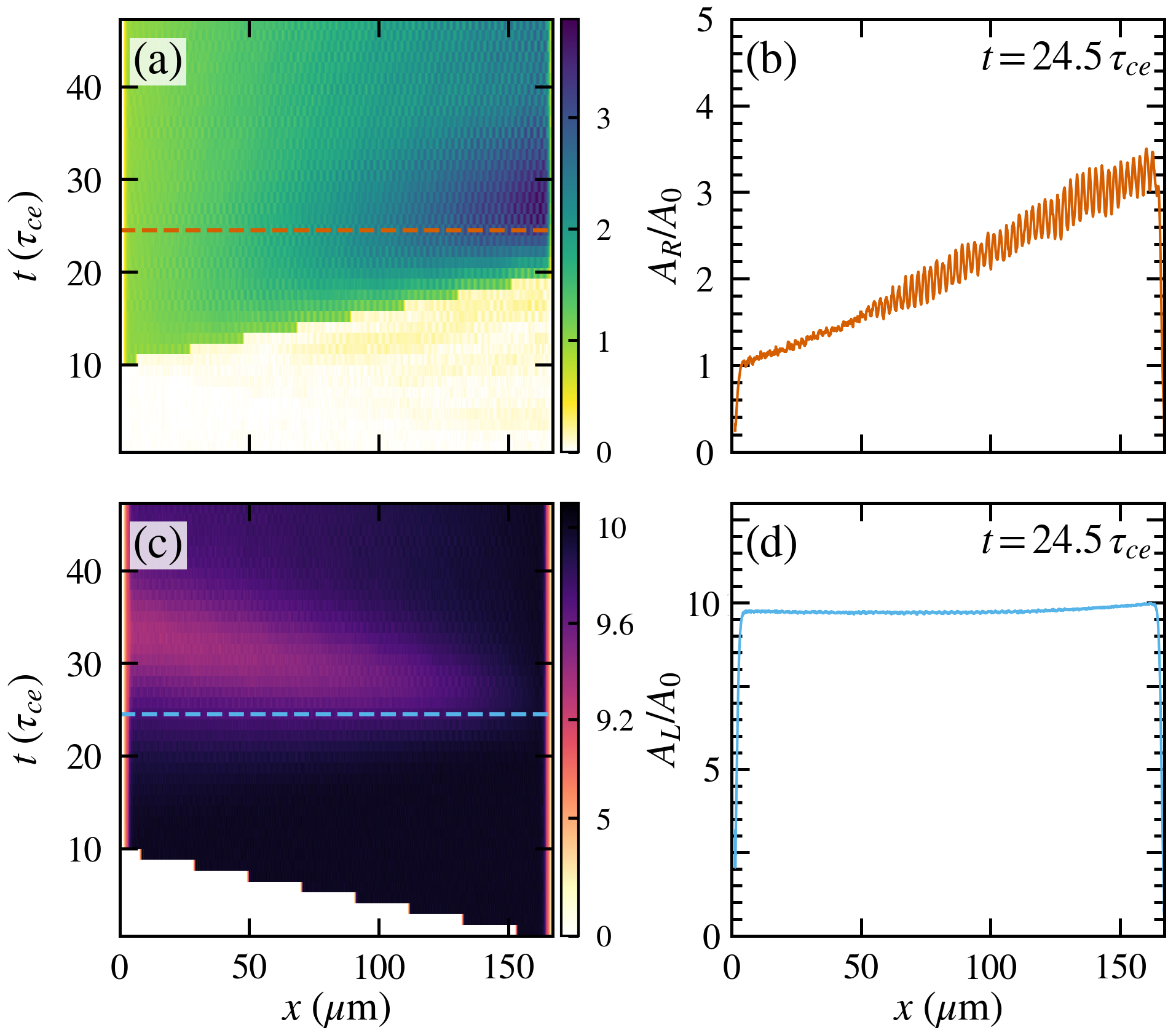} 
\end{center}
\caption{Snapshots of right-moving seed envelope $A_R$ (top) and left-moving pump envelope $A_L$ (bottom) in an example PIC simulation with $\lambda_1 = 409.8$ nm and $B_0=600$ T (cyclotron period $\tau_{\text{ce}}\approx 59.5$ fs). 
The lineouts at $t=24.5 \tau_{\text{ce}}$ shows the near exponential growth of the seed (b) and a slight depletion of the pump (d).
All envelopes are normalized to the incident seed amplitude ($A_0$).} 
\label{fig: envfigs}
\end{figure}

We measure the convective gain of the seed laser from PIC data in a quasi-steady state before pump depletion occurs, as shown in Fig.~\ref{fig: envfigs} for the example $B_0=600$ T and $\lambda_1 = 409.8$ nm. 
At each $B_0$, we scan $\lambda_1$ in an ensemble of simulations with a step size \mbox{$\Delta \lambda_1 = 0.3$ nm} to construct a numerical gain spectrum. 
Due to phase-matching conditions of three-wave interactions $\omega_2 = \omega_0-\omega_1$ and $k_2=k_0-k_1$, scanning the vacuum wavelength $\lambda_1=2\pi c/\omega_1$ is equivalent to scanning the mediating plasma wave frequency $\omega_2$.

\begin{figure} [b]
\begin{center}
\includegraphics[scale=1.0]{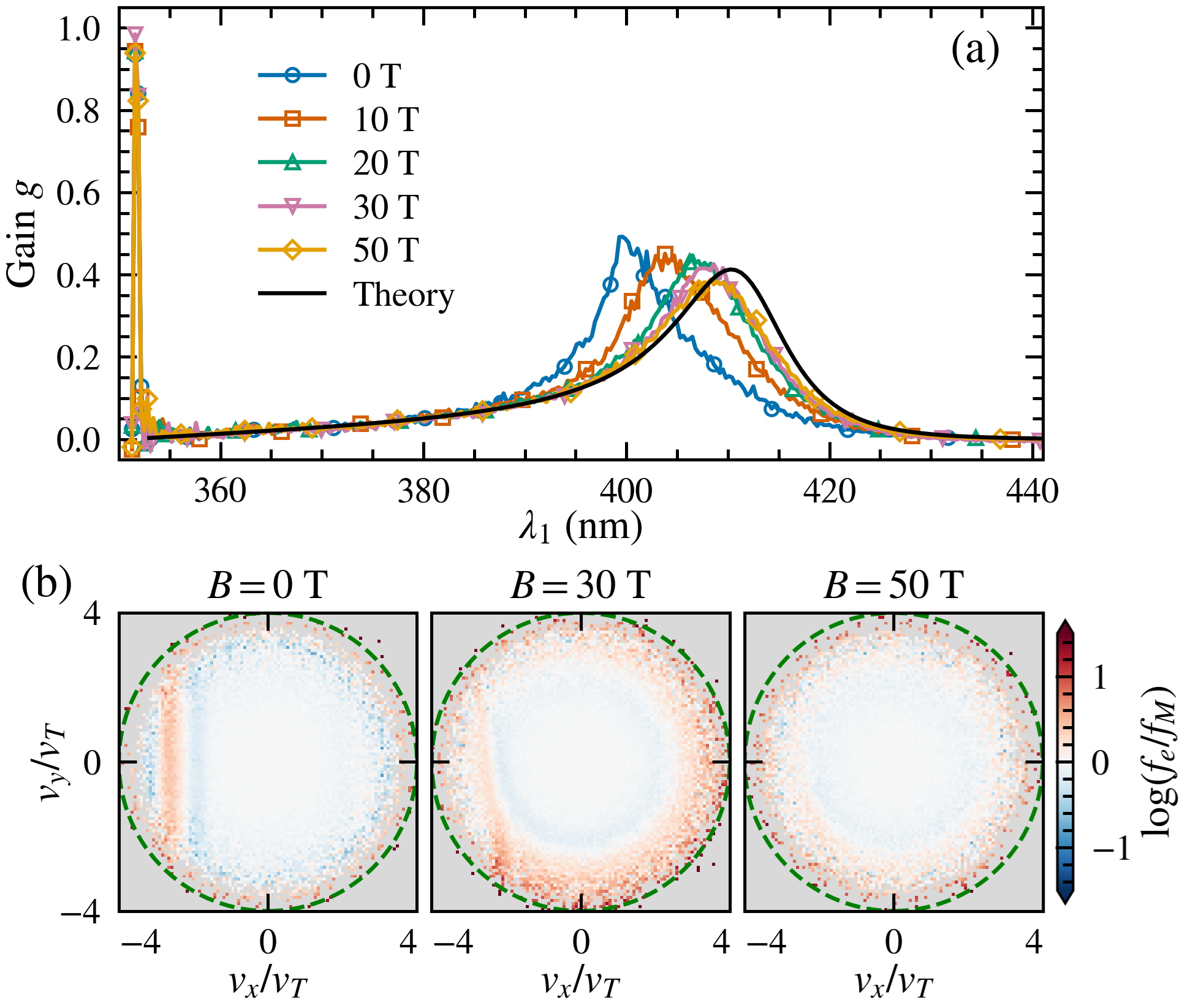} 
\end{center}
\caption{(a) Gain spectrum in quasi-steady state ($\theta=90^\circ$). The narrow peak near $\lambda_1=351$ nm is from Brillouin scattering and the broad peak is from Raman scattering. Compared to the unmagnetized linear theory (black), particle trapping induces nonlinear frequency shift for $B_0\lesssim 30$ T. In stronger $B_0$, cyclotron motion detraps particles, restoring the validity of the linear theory. 
(b) Electron distribution functions $f_e(v_x, v_y)$, normalized to the initial Maxwellian $f_M$ and the thermal speed $v_T$, when $\lambda_1$ is near the spectral peak.
When $B_0=0$, a plateau develops near the phase velocity of the left-moving plasma wave. When $B_0=30$ T, cyclotron motion rotates the plateau and surfatron accelerates electrons to high speed. At $B_0=50$ T, trapping is disrupted and a plateau never forms.} 
\label{fig: lowBfigs}
\end{figure}

\textit{Results.} 
In the $B_0=0$ limit, the linear gain calculated from the magnetized Vlasov theory \cite{MoodyLinTheory} is the same as the unmagnetized theory \cite{PUPIL}, which is shown by the black line in Fig.~2(a). The linear gain differs from the PIC gain (colored lines) due to the nonlinear frequency downshift of the mediating plasma wave \cite{morales1972nonlinear}.
The frequency shift is caused by particle trapping \cite{yin2007stimulated} that modifies the electron distribution function $f_e$, which develops islands in the phase space.  
In the $v_x$-$v_y$ plane, the phase-space islands manifest as a plateau when $v_x=\omega_2/k_2\approx-2.7 v_T$ is near the phase velocity of the left-propagating plasma wave, as shown in Fig.~2(b), where $v_T$ is the thermal speed.
Flattening of the distribution function reduces collisionless damping and leads to kinetic inflation in the deeply nonlinear regime \cite{rose2001kinetic}.

In weak fields \mbox{$B_0 \lesssim 50$ T}, the linear theory gain is barely changed, whereas nonlinear trapping is noticeably suppressed, which brings the Raman peak closer to linear theory. 
The competition between cyclotron motion and the bounce motion in the wave potential well leads to detrapping of resonant particles. When the particles eventually detrap, they are accelerated to high speed in the $v_\perp$ plane, a process known as surfatron acceleration \cite{sagdeev1973influence,Dawson1983}. 
The spatially averaged distribution functions are shown in Fig.~2(b) for seed wavelength $\lambda_1$ such that the driven plasma wave frequency is $\omega_2=\omega_{\text{UH}} = (\omega_{\text{ce}}^2+\omega_{\text{pe}}^2)^{1/2}$, namely, the cold upper-hybrid frequency.
Near \mbox{$B_0=30$ T}, wave trapping and cyclotron detrapping roughly balance. A plateau in $f_e(v_x, v_y)$ still develops, but surfatron acceleration sweeps trapped particles away, forming a high-energy tail that is rotationally symmetric in the $v_\perp$ plane.
At \mbox{$B_0=50$ T}, cyclotron detrapping dominates, and a plateau never forms. The distribution function instead broadens to have a non-Maxwellian tail and collisionless damping remains strong, suppressing Raman scattering and raising the threshold of kinetic inflation \cite{Winjum18,Lee26}.
A heuristic threshold $B_c$ beyond which detrapping dominates can be estimated as follows. 
Assuming an electrostatic wave $\mathbf{E}=-\Phi_0\nabla \cos(kx - \omega_{\text{UH}}t)$, the equation of motion of a trapped electron in the wave frame ($x'$) with phase velocity $v_p = \omega_{\text{UH}}/k$ is $\ddot{x'} = -\frac{\omega_b^2}{k}\sin(k x') - \omega_{\text{ce}} \dot{y}$ and $\ddot{y} = \omega_{\text{ce}}\dot{x}$, where  $\omega_b^2=ek^2\Phi_0/m_e$ is the bounce frequency.
A trapped electron escapes when the transverse velocity $v_{\times}=E_x/B_0=\omega_b^2/(k\omega_{\text{ce}})$ is large enough for $E_x + \dot{y}B_0$ to vanish.
Denoting $V\equiv v_{\times}/v_p = \omega_b^2/(\omega_{\text{ce}}\omega_{\text{UH}})$, cyclotron motion dominates when $V\ll 1$, where the net work done by the wave on the electron averages to zero.
On the other hand, when $V\gtrsim 1$, the average work over a gyro period no longer vanishes, so the electron surfatron accelerates. The detrapping condition $V\sim1$ is only heuristic, because if the electron starts at different locations in the potential well or has a sufficient $\dot{x}$, then it can escape with $\dot{y}$ much smaller than $v_\times$ \cite{Winjum18}. 
Nevertheless, for our plasma parameters, $V\sim1$ gives a threshold \mbox{$B_c\approx27$ T}, consistent with the PIC results.

\begin{figure} [b]
\begin{center}
\includegraphics[scale=1]{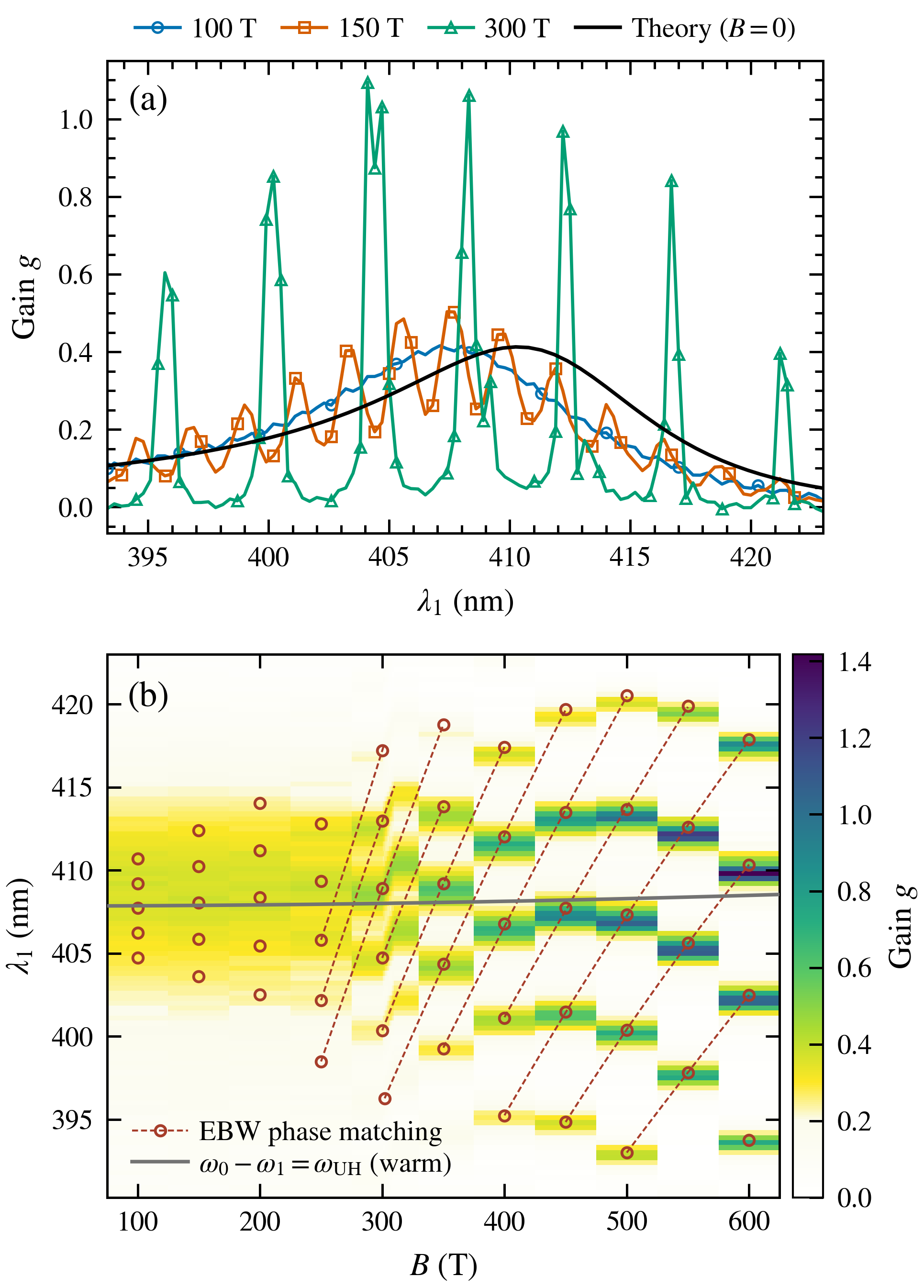} 
\end{center}
\caption{Transition from Raman to Bernstein scattering ($\theta=90^\circ$) in PIC simulations. 
(a) The Raman peak splits into resonances mediated by electron Bernstein waves (EBWs), whose frequency spacing is approximately $\omega_{\text{ce}}$. At stronger $B_0$, the EBW resonances grow stronger and narrower. Between EBW peaks, the gain decreases to zero, where Raman scattering ceases to exist. Simulations use 1000 particles per cell (PPC).
(b) The gain spectra (color) is explained by phase-matching with EBWs (circles). The scan is finer near $B_0=300$ T. The resonances are strongest near the upper-hybrid frequency $\omega_{\text{UH}}$ (solid). Simulations use 100 PPC.
} 
\label{fig: highBfigs}
\end{figure}

In stronger magnetic fields, the Raman resonance is gradually replaced by resonances mediated by EBWs, as the PIC results show in Fig.~3 where $\mathbf{B}$ is exactly perpendicular to the wave propagation direction ($\theta=90^\circ$). 
At \mbox{$B_0=100$ T}, wiggles begin to develop on the broad Raman peak. 
As $B_0$ increases, the wiggles are further separated and their contrast increases. 
By \mbox{$B_0=150$ T}, the wiggles have emerged as well separated peaks. With further increase of $B_0$, the widths of the peaks become smaller than their separations, and the heights of the peaks far exceed the Raman peak. Interestingly, between the peaks, the gain decreases to zero, as shown in the example \mbox{$B_0=300$ T}, which means that instead of coexisting with Raman, the EBW resonances replace Raman. 
For Fig.~3(a), to resolve this transition from trapping-dominated motion to cyclotron-dominated motion, we use 1000 particles-per-cell (PPC) to achieve numerical convergence. Away from this transition, at either higher or lower $B_0$, the PPC required to achieve convergence is significantly lower.
For Fig.~3(b), we use PPC=100 and show the gain spectra (color) for a coarse scan of \mbox{$\Delta B_0=50$ T} and a fine scan of \mbox{$\Delta B_0=1$ T} in the range \mbox{300 T}$\le B_0\le$\mbox{310 T}. 
The circles are analytically expected locations of EBW resonances, obtained 
as numerical roots of $\omega_0=\omega_1(k_0-k_2) + \omega_2(k_2)$
where $\omega(k)$ satisfies the linear dispersion relation \cite{bernstein1958waves}, including finite-$k$ effects. 
When $k_2\rightarrow\infty$, the frequencies $\omega_2\rightarrow n\omega_{\text{ce}}$, where $n$ are integers. 
The dashed lines in Fig.~3(b) connect resonances with the same $n$. 
For a given $n$, because $\omega_{\text{ce}}$ increases with $B_0$, the frequencies of the mediating wave increase, leading to increases in resonant seed wavelengths.  
This smooth increase is resolved with the fine scan \mbox{$\Delta B_0=1$ T} near \mbox{$B_0=300$ T}. 
Phase matching confirms that the resonances are mediated by EBWs.

The finite widths of the EBW resonances show that individual Bernstein waves are damped in collisionless PIC simulations even for exact perpendicular propagation, a phenomenon not captured by linear theory. In linear theory, where unperturbed trajectories are circular gyro orbits, the collisionless damping rate of Bernstein waves is exactly zero at $\theta=90^\circ$ \cite{stix1992waves}. The vanishing collisionless damping rate of individual EBWs at $B_0>0$, in contradiction to a finite Landau damping rate at $B_0\rightarrow 0$, is known as the Bernstein-Landau paradox \cite{tataronis1970cyclotron,Sukhorukov,Valentini04,Charles2021}, which is often studied as an initial value problem, without regard to how the initial value is established. 
In our simulations, EBWs are excited by counter propagating lasers. The three-wave interaction injects energy into the mediating wave, which undergoes collisionless damping that gives rise to the finite width of the resonance. 
As $B_0$ decreases, EBW resonances merge into a single broad Raman resonance when $\omega_{\text{ce}}$ becomes comparable to the damping rate $\nu$, which is finite.
Interestingly, even though $\omega_{\text{UH}}$ is not an eigenmode of kinetic plasmas, the EBW resonances are strongest when their frequencies are near  $\omega_{\text{UH}}$, which is shown as the gray line in Fig.~3(b).

\begin{figure} [ht]
\begin{center}
\includegraphics[scale=1]{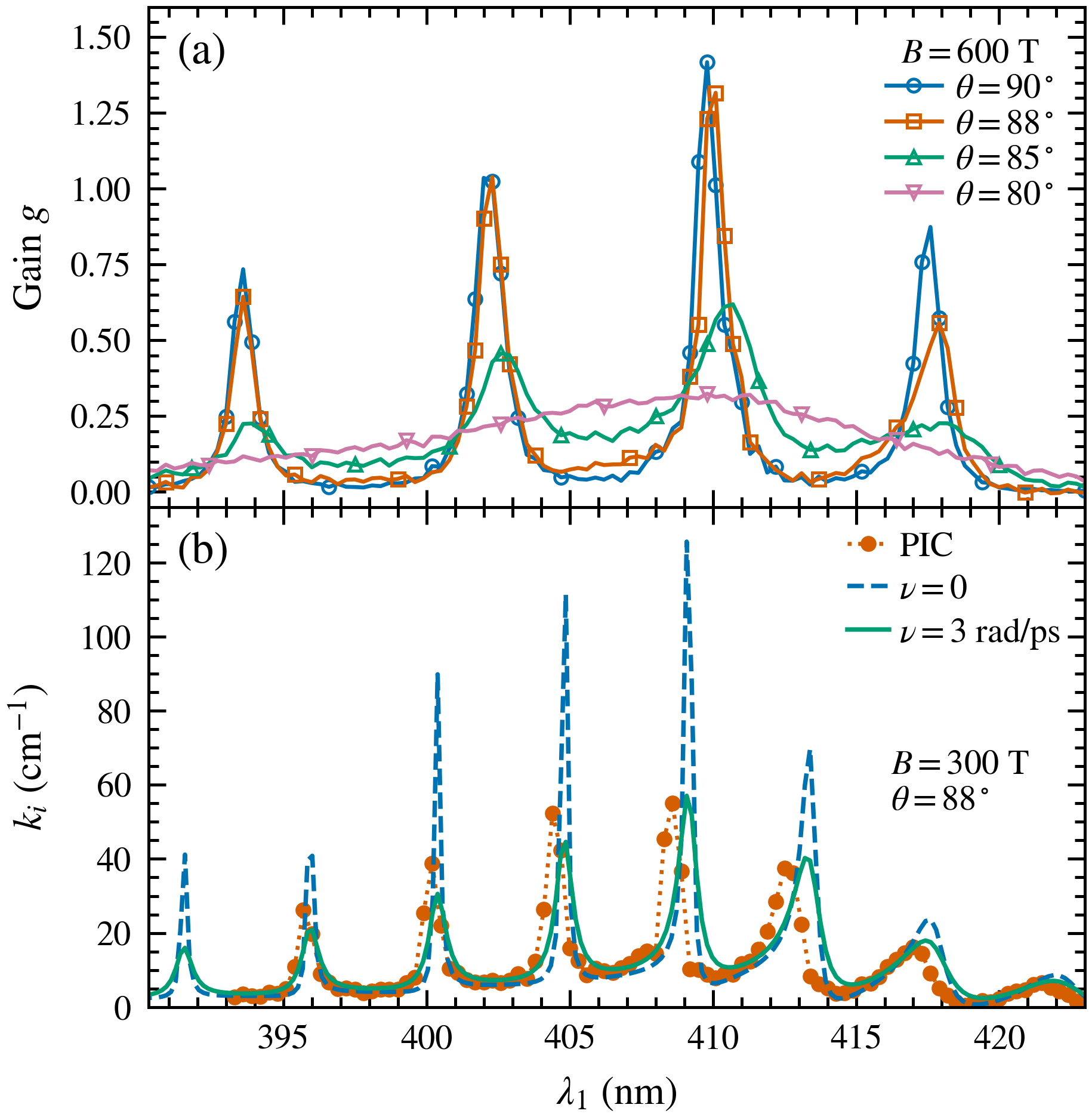} 
\end{center}
\caption{Transition from Raman to Bernstein scattering is sensitive to propagation angle. (a) As $\theta$ decreases away from perpendicular, the EBW resonances become wider and weaker, and they merge back into a single broad Raman peak when $\theta\gtrsim 80^\circ$ at \mbox{$B_0 = 600$ T}.  
(b) Linear Vlasov theory, which has zero collisionless damping at $\theta=90^\circ$, can match PIC data only when collisional damping ($\nu$) is artificially added, as shown in the example $B_0=300$ T and $\theta=88^\circ$.
}
\label{fig: anglefig}
\end{figure}

As $\theta$ decreases, the increase of linear damping merges EBWs back into a single broad Raman peak, so Bernstein scattering is observable only when the mediating plasma waves propagate near perpendicular to the magnetic field. In linear theory, the damping rate is proportional to the imaginary part of the susceptibility Im$(\chi_{xx}) \propto \exp [(\omega-n\omega_{\text{ce}})/k_\parallel v_T]$, which becomes nonzero at finite $k_\parallel = k\cos\theta$. 
Heuristically, when $k_\parallel v_T\gtrsim \omega_{\text{ce}}$, the widths of individual EBWs become larger than their spacing. Estimating the wave vector by that of the Langmuir wave $k\simeq \omega_{\text{pe}}/v_T$ and using small angle approximation $\sin\alpha\simeq\alpha = 90^\circ-\theta$, the condition that EBWs merge can be estimated as $\alpha> \alpha_c \simeq\omega_{\text{ce}}/\omega_{\text{pe}}$. 
Under our plasma conditions and for \mbox{$B_0=600$ T}, the estimate gives $\alpha_c\approx 10^\circ$, which agrees well with PIC results shown in Fig.~4(a), where EBWs completely merge by $\theta=80^\circ$.

Linear theory underpredicts the damping rate at small $\alpha$, as shown in Fig.~4(b) where $\theta=88^\circ$ and \mbox{$B_0=300$ T}. Compared to the gain spectrum of PIC (dots), the Vlasov theory (dashed) predicts weaker collisionless damping, so the EBW peaks are stronger and narrower. 
Adding an artificial collisional damping \mbox{$\nu=3$ rad/ps} (solid) that modifies the EBW poles to $\omega-k_\parallel v_\parallel -n \omega_{\text{ce}} + i\nu$ allows a better matching with the PIC data. 
The added damping rate is comparable to the physical collision rate under our plasma conditions, but notice that the PIC simulations have collisions turned off. 
The locations of the peaks are downshifted in PIC, possibly due to particle trapping, similar to Fig.~2(a).

\textit{Discussion.} The sensitivity of the Raman-Bernstein transition to the plasma wave propagation angle allows for using laser scattering to measure local magnetic field vectors. 
The properties of EBWs, revealed in this Letter for laser backscattering, are also applicable to side scatterings. 
Using spontaneous laser Thomson scattering \cite{froula2010plasma,sitenko2012electromagnetic} or stimulated cross-beam energy transfer (CBET) \cite{Longman25,Turnbull26}, 
where the pump ($\mathbf{k}_0$) and seed ($\mathbf{k}_1$) propagate at different angles, and therefore overlap only in a spatially localized region, EBW scattering allows measuring the component of $\mathbf{B}$ that is perpendicular to $\Delta\mathbf{k}=\mathbf{k}_0-\mathbf{k_1}$. The magnitude of $B_\perp$ is inferrable from the separation of EBW peaks, which is insensitive to the plasma density and temperature. 
This measurement technique is suitable for $B_\perp \gtrsim 100$ T, a regime that is particularly challenging to diagnose for existing path-integrated techniques, such as proton and electron radiography. EBW scattering may thus provide useful diagnostics for magnetized high-energy-density science (HEDS) experiments.

The implications of EBW resonances for magnetized HEDS remain subjects of future work. 
In particular, the gain away from EBW resonances is suppressed to zero, while the gain on EBW resonances can far exceed the Raman gain for $B_\perp$ of a few MG, which can be generated by internal plasma dynamics or by flux compression of externally imposed seed fields. 
As $B_\perp$ increases, a broad but weak Raman resonance is replaced by a frequency comb like spectrum of narrow but strong resonances. For nonlinear plasma optics, the frequency comb may allow plasma-mediated laser amplification to generate ultrashort pulses.
For inertial confinement fusion (ICF), how the frequency comb affects the overall nonlinear reflectivity and the effectiveness of bandwidth mitigation of LPIs are open questions. 
Our results focus on how EBWs replace the Langmuir wave. Analogously, ion Bernstein waves may have similar effects on replacing the ion-acoustic wave, which our simulations do not resolve. 
If the drive lasers in ICF can be detuned to avoid Bernstein resonances, then Raman scattering, Brillouin scattering, and CBET may be strongly suppressed for $\Delta\mathbf{k} \perp \mathbf{B}$. 
Finally, our discovery that the collisionless damping rate of individual Bernstein wave branches is nonzero even for $\theta=90^\circ$, in contrast to what established linear theory predicts, also raises questions in fundamental plasma wave theory of what analytical technique is appropriate when the bounce frequency $\omega_b$ is larger than the cyclotron frequency $\omega_{\text{ce}}$. Effects of gradients, polarization rotation, and non-collinear side scattering also remain to be investigated.

In summary, we use PIC simulations to show that the long-theorized Bernstein scattering occurs in parameter regimes relevant for laser-driven HEDS experiments. We show in weak magnetic fields, electron trapping is suppressed, so the nonlinear stage of magnetized Raman scattering behaves closer to what linear theory expects when the bounce frequency $\omega_b\lesssim (\omega_{\text{ce}}\omega_{\text{UH}})^{1/2}$. 
Moreover, we show that as $B_\perp$ increases, the Langmuir wave mediated Raman resonance is gradually replaced by EBWs mediated resonances that are much stronger than narrower, while the coupling is suppressed between EBW resonances. As the propagation angle decreases by $\alpha\gtrsim \omega_{\text{ce}}/\omega_{\text{pe}}$ away from exact perpendicular, the EBW resonances merge back into a broad Raman-like resonance. 
Regardless of the propagation angle, linear theory underpredicts collisionless damping of EBWs, and the damping rate of the individual EBW branch is nonzero even at exact perpendicular propagation. 
Magnetized Vlasov theory with added damping can capture the EBW gain spectrum up to a small nonlinear frequency downshift. 
Our results open new avenues for field diagnostics, plasma photonics, and LPIs mitigation.

\begin{acknowledgments}
This work was supported by U.S. Department of Energy under award number DE-SC0026586. The work performed by Lawrence Livermore National Laboratory was under the auspices of the U.S. Department of Energy under Contract DE-AC52-07NA27344.
The EPOCH code was in part funded by the UK EPSRC grants EP/G054950/1, EP/G056803/1, EP/G055165/1 and EP/ M022463/1.
\end{acknowledgments}


\bibliography{EBW}

@article{Valentini04,
    author = {Valentini, F. and Veltri, P. and Mangeney, André},
    title = {Nonlinear evolution of high frequency electrostatic waves in a magnetized plasma: {Bernstein‐Landau} paradox revisited},
    journal = {AIP Conference Proceedings},
    volume = {703},
    number = {1},
    pages = {80-83},
    year = {2004},
    month = {04},
    issn = {0094-243X},
    doi = {10.1063/1.1718438},
    fullurl = {https://doi.org/10.1063/1.1718438},
    fulleprint = {https://pubs.aip.org/aip/acp/article-pdf/703/1/80/11508328/80_1_online.pdf},
}

@article{Charles2021,
  author  = {Charles, Fr\'ed\'erique and Despr\'es, Bruno and Rege, Alexandre and Weder, Ricardo},
  title   = {The Magnetized {Vlasov--Amp\`ere} System and the {Bernstein--Landau} Paradox},
  journal = {Journal of Statistical Physics},
  year    = {2021},
  volume  = {183},
  number  = {2},
  pages   = {23},
  doi     = {10.1007/s10955-021-02755-9},
  url     = {https://doi.org/10.1007/s10955-021-02755-9},
  issn    = {1572-9613}
}

@article{Bott17, 
title={Proton imaging of stochastic magnetic fields}, 
volume={83}, 
DOI={10.1017/S0022377817000939}, 
number={6}, 
journal={Journal of Plasma Physics}, 
author={Bott, A. F. A. and Graziani, C. and Tzeferacos, P. and White, T. G. and Lamb, D. Q. and Gregori, G. and Schekochihin, A. A.}, 
year={2017}, 
pages={905830614}
}

@article{Zhang22,
author = {Chaojie Zhang  and Yipeng Wu  and Mitchell Sinclair  and Audrey Farrell  and Kenneth A. Marsh  and Irina Petrushina  and Navid Vafaei-Najafabadi  and Apurva Gaikwad  and Rotem Kupfer  and Karl Kusche  and Mikhail Fedurin  and Igor Pogorelsky  and Mikhail Polyanskiy  and Chen-Kang Huang  and Jianfei Hua  and Wei Lu  and Warren B. Mori  and Chan Joshi },
title = {Mapping the self-generated magnetic fields due to thermal {Weibel} instability},
journal = {Proceedings of the National Academy of Sciences},
volume = {119},
number = {50},
pages = {e2211713119},
year = {2022},
doi = {10.1073/pnas.2211713119},
fullURL = {https://www.pnas.org/doi/abs/10.1073/pnas.2211713119},
fulleprint = {https://www.pnas.org/doi/pdf/10.1073/pnas.2211713119},
}

@article{sagdeev1973influence,
  title={Influence of transverse magnetic field on Landau damping},
  author={Sagdeev, RZ and Shapiro, VD},
  journal={Soviet Journal of Experimental and Theoretical Physics Letters},
  volume={17},
  pages={279},
  year={1973}
}

@article{Shi17three,
  title = {Three-wave scattering in magnetized plasmas: From cold fluid to quantized Lagrangian},
  author = {Shi, Yuan and Qin, Hong and Fisch, Nathaniel J.},
  journal = {Phys. Rev. E},
  volume = {96},
  issue = {2},
  pages = {023204},
  numpages = {25},
  year = {2017},
  month = {Aug},
  publisher = {American Physical Society},
  doi = {10.1103/PhysRevE.96.023204},
  fullurl = {https://link.aps.org/doi/10.1103/PhysRevE.96.023204}
}

@article{Jia17,
    author = {Jia, Qing and Shi, Yuan and Qin, Hong and Fisch, Nathaniel J.},
    title = {Kinetic simulations of laser parametric amplification in magnetized plasmas},
    journal = {Physics of Plasmas},
    volume = {24},
    number = {9},
    pages = {093103},
    year = {2017},
    month = {08},
    issn = {1070-664X},
    doi = {10.1063/1.4998168},
    fullurl = {https://doi.org/10.1063/1.4998168},
    fulleprint = {https://pubs.aip.org/aip/pop/article-pdf/doi/10.1063/1.4998168/15657914/093103_1_online.pdf},
}

@article{Schumaker13,
  title = {Ultrafast Electron Radiography of Magnetic Fields in High-Intensity Laser-Solid Interactions},
  author = {Schumaker, W. and Nakanii, N. and McGuffey, C. and Zulick, C. and Chyvkov, V. and Dollar, F. and Habara, H. and Kalintchenko, G. and Maksimchuk, A. and Tanaka, K. A. and Thomas, A. G. R. and Yanovsky, V. and Krushelnick, K.},
  journal = {Phys. Rev. Lett.},
  volume = {110},
  issue = {1},
  pages = {015003},
  numpages = {5},
  year = {2013},
  month = {Jan},
  publisher = {American Physical Society},
  doi = {10.1103/PhysRevLett.110.015003},
  fullurl = {https://link.aps.org/doi/10.1103/PhysRevLett.110.015003}
}

@article{Schaeffer23,
  title = {Proton imaging of high-energy-density laboratory plasmas},
  author = {Schaeffer, Derek B. and Bott, Archie F. A. and Borghesi, Marco and Flippo, Kirk A. and Fox, William and Fuchs, Julien and Li, Chikang and S{\'e}guin, Fredrick H. and Park, Hye-Sook and Tzeferacos, Petros and Willingale, Louise},
  journal = {Rev. Mod. Phys.},
  volume = {95},
  issue = {4},
  pages = {045007},
  numpages = {48},
  year = {2023},
  month = {Dec},
  publisher = {American Physical Society},
  doi = {10.1103/RevModPhys.95.045007},
  url = {https://link.aps.org/doi/10.1103/RevModPhys.95.045007}
}

@inproceedings{deeney2022advances,
  title={Advances in Direct-Drive Fusion, High-Energy-Density Physics, and Laser Technologies at the Laboratory for Laser Energetics},
  author={Deeney, C},
  booktitle={2022 IEEE International Conference on Plasma Science (ICOPS)},
  pages={1--2},
  year={2022},
  organization={IEEE}
}

@article{Kang_2025,
title={Effects of moderate bandwidth on laser absorption and stimulated {Raman} sidescattering},
author={Kang, Ning and Zhao, Yao and Lei, Anle and Liu, Huiya and Li, Fujian and Xu, Guoxiao and An, Honghai and Xiong, Jun and Wang, Ruirong and Xie, Zhiyong and others},
journal={Nuclear Fusion},
doi = {10.1088/1741-4326/ada629},
fullurl = {https://doi.org/10.1088/1741-4326/ada629},
year = {2025},
month = {jan},
publisher = {IOP Publishing},
volume = {65},
number = {2},
pages = {026042},
}

@article{Shi17compression,
  title = {Laser-pulse compression using magnetized plasmas},
  author = {Shi, Yuan and Qin, Hong and Fisch, Nathaniel J.},
  journal = {Phys. Rev. E},
  volume = {95},
  issue = {2},
  pages = {023211},
  numpages = {5},
  year = {2017},
  month = {Feb},
  publisher = {American Physical Society},
  doi = {10.1103/PhysRevE.95.023211},
  fullurl = {https://link.aps.org/doi/10.1103/PhysRevE.95.023211}
}

@article{Edwards19,
  title = {Laser Amplification in Strongly Magnetized Plasma},
  author = {Edwards, Matthew R. and Shi, Yuan and Mikhailova, Julia M. and Fisch, Nathaniel J.},
  journal = {Phys. Rev. Lett.},
  volume = {123},
  issue = {2},
  pages = {025001},
  numpages = {7},
  year = {2019},
  month = {Jul},
  publisher = {American Physical Society},
  doi = {10.1103/PhysRevLett.123.025001},
  fullurl = {https://link.aps.org/doi/10.1103/PhysRevLett.123.025001}
}

@article{Manzo22,
    author = {Manzo, Lili and Edwards, Matthew R. and Shi, Yuan},
    title = {Enhanced collisionless laser absorption in strongly magnetized plasmas},
    journal = {Physics of Plasmas},
    volume = {29},
    number = {11},
    pages = {112704},
    year = {2022},
    month = {11},
    issn = {1070-664X},
    doi = {10.1063/5.0100727},
    fullurl = {https://doi.org/10.1063/5.0100727},
    fulleprint = {https://pubs.aip.org/aip/pop/article-pdf/doi/10.1063/5.0100727/16629260/112704_1_online.pdf},
}

@article{Brodin13, 
title={Stimulated {Brillouin} scattering in magnetized plasmas}, 
volume={79}, 
DOI={10.1017/S0022377813000664}, 
number={6}, 
journal={Journal of Plasma Physics}, 
author={Brodin, G. and Stenflo, L.}, 
year={2013}, 
pages={983–986}
}

@article{Karmendy79,
title = {{Brillouin} scattering from $\text{CO}_2$ laser plasmas in strong transverse magnetic fields},
journal = {Journal of Magnetism and Magnetic Materials},
volume = {11},
number = {1},
pages = {379-382},
year = {1979},
issn = {0304-8853},
doi = {https://doi.org/10.1016/0304-8853(79)90294-4},
fullurl = {https://www.sciencedirect.com/science/article/pii/0304885379902944},
author = {C. Karmendy and W. Halverson and B. Lax}
}

@article{Shi19,
  title = {Three-wave interactions in magnetized warm-fluid plasmas: General theory with evaluable coupling coefficient},
  author = {Shi, Yuan},
  journal = {Phys. Rev. E},
  volume = {99},
  issue = {6},
  pages = {063212},
  numpages = {17},
  year = {2019},
  month = {Jun},
  publisher = {American Physical Society},
  doi = {10.1103/PhysRevE.99.063212},
  fullurl = {https://link.aps.org/doi/10.1103/PhysRevE.99.063212}
}

@article{Grebogi80,
    author = {Grebogi, Celso and Liu, C. S.},
    title = {{Brillouin and Raman} scattering of an extraordinary mode in a magnetized plasma},
    journal = {The Physics of Fluids},
    volume = {23},
    number = {7},
    pages = {1330-1335},
    year = {1980},
    month = {07},
    issn = {0031-9171},
    doi = {10.1063/1.863146},
    fullurl = {https://doi.org/10.1063/1.863146},
    fulleprint = {https://pubs.aip.org/aip/pfl/article-pdf/23/7/1330/12266483/1330_1_online.pdf},
}

@article{Li23TPD,
  title = {Anomalous staged hot-electron acceleration by two-plasmon decay instability in magnetized plasmas},
  author = {Li, X. X. and Cheng, R. J. and Wang, Qing and Liu, D. J. and Lv, S. Y. and Huang, Z. M. and Zhang, S. T. and Li, X. M. and Chen, Z. J. and Wang, Qiang and Liu, Z. J. and Cao, L. H. and Zheng, C. Y. and He, X. T.},
  journal = {Phys. Rev. E},
  volume = {108},
  issue = {5},
  pages = {L053201},
  numpages = {6},
  year = {2023},
  month = {Nov},
  publisher = {American Physical Society},
  doi = {10.1103/PhysRevE.108.L053201},
  fullurl = {https://link.aps.org/doi/10.1103/PhysRevE.108.L053201}
}

@article{Barr84,
    author = {Barr, H. C. and Boyd, T. J. M. and Gardner, L. R. T. and Rankin, R.},
    title = {{Raman} and two‐plasmon decay instabilities in a magnetized plasma},
    journal = {The Physics of Fluids},
    volume = {27},
    number = {11},
    pages = {2730-2737},
    year = {1984},
    month = {11},
    issn = {0031-9171},
    doi = {10.1063/1.864577},
    fullurl = {https://doi.org/10.1063/1.864577},
    fulleprint = {https://pubs.aip.org/aip/pfl/article-pdf/27/11/2730/12729435/2730_1_online.pdf},
}

@Article{Shi26,
AUTHOR = {Shi, Yuan and Moody, John D.},
TITLE = {Particle-in-Cell Simulations of Laser Crossbeam Energy Transfer via Magnetized Ion-Acoustic Wave},
JOURNAL = {Physics},
VOLUME = {8},
YEAR = {2026},
NUMBER = {1},
ARTICLE-NUMBER = {25},
fullurl = {https://www.mdpi.com/2624-8174/8/1/25},
ISSN = {2624-8174},
DOI = {10.3390/physics8010025}
}

@article{Nabil98,
doi = {10.1088/0031-8949/57/2/018},
fullurl = {https://doi.org/10.1088/0031-8949/57/2/018},
year = {1998},
month = {feb},
publisher = {},
volume = {57},
number = {2},
pages = {253},
author = {Nabil M Laham and Ali S Al Nasser and Ahmed M Khateeb},
title = {Effects of Axial Magnetic Fields on Backward {Raman}
Scattering in Inhomogeneous Plasmas},
journal = {Physica Scripta},
}

@article{Platzman68,
  title = {Light Scattering from a Plasma in a Magnetic Field},
  author = {Platzman, P. M. and Wolff, P. A. and Tzoar, N.},
  journal = {Phys. Rev.},
  volume = {174},
  issue = {2},
  pages = {489--494},
  numpages = {0},
  year = {1968},
  month = {Oct},
  publisher = {American Physical Society},
  doi = {10.1103/PhysRev.174.489},
  fullurl = {https://link.aps.org/doi/10.1103/PhysRev.174.489}
}

@article{Liu1986parametric,
  title={Parametric instabilities in a magnetized plasma},
  author={Liu, Chuan Sheng and Tripathi, VK},
  journal={Physics Reports},
  volume={130},
  number={3},
  pages={143--216},
  year={1986},
  publisher={Elsevier}
}

@article{Stenflo81,
  title = {Comments on stimulated scattering of electromagnetic waves by electron {Bernstein} modes in a plasma},
  author = {Stenflo, L.},
  journal = {Phys. Rev. A},
  volume = {23},
  issue = {5},
  pages = {2730--2731},
  numpages = {0},
  year = {1981},
  month = {May},
  publisher = {American Physical Society},
  doi = {10.1103/PhysRevA.23.2730},
  fullurl = {https://link.aps.org/doi/10.1103/PhysRevA.23.2730}
}

@article{Simon95,
    author = {Simon, Albert},
    title = {Parametric excitation of {Bernstein} modes in laser‐produced plasma},
    journal = {Physics of Plasmas},
    volume = {2},
    number = {10},
    pages = {3832-3843},
    year = {1995},
    month = {10},
    issn = {1070-664X},
    doi = {10.1063/1.871082},
    fullurl = {https://doi.org/10.1063/1.871082},
    fulleprint = {https://pubs.aip.org/aip/pop/article-pdf/2/10/3832/19118839/3832_1_online.pdf},
}

@article{Sharma80,
  title = {Nonlinear scattering of upper hybrid laser radiation by electron {Bernstein} modes in a plasma},
  author = {Sharma, R. R. and Salimullah, Md. and Tripathi, V. K.},
  journal = {Phys. Rev. A},
  volume = {21},
  issue = {3},
  pages = {942--947},
  numpages = {0},
  year = {1980},
  month = {Mar},
  publisher = {American Physical Society},
  doi = {10.1103/PhysRevA.21.942},
  url = {https://link.aps.org/doi/10.1103/PhysRevA.21.942}
}

@article{Paknezhad16, 
title={Effect of dc-magnetic field on the growth rate of {Raman} backscattering of {X-mode} laser in magnetized collisional plasma}, 
volume={34}, 
DOI={10.1017/S0263034615000932}, 
number={1}, 
journal={Laser and Particle Beams}, 
author={Paknezhad, Alireza}, 
year={2016}, 
pages={80–85}
}

@article{Lee26, 
title={Effect of small magnetic fields on stimulated {Raman} scattering in the kinetic regime}, 
volume={92}, 
DOI={10.1017/S0022377826102384}, 
number={5}, 
journal={Journal of Plasma Physics}, 
author={Lee, Roman P. and Winjum, Benjamin and Spencer, S.J. and Bolaños, Simon and Bailly-Grandvaux, Mathieu and Manuel, Mario J.-E. and Tsung, Frank S. and Beg, Farhat and Mori, Warren B.}, 
year={2026}, 
pages={E133}
}

@article{Bolanos25,
    author = {Bolaños, S. and Bailly-Grandvaux, M. and Winjum, B. J. and Manuel, M. J.-E. and Bogale, A. and Filkins, T. and Froula, D. H. and Lee, R. and Mori, W. and Tsung, F. and Walsh, C. A. and Beg, F. N.},
    title = {Direct evidence of the effect of a moderate external magnetic field on stimulated {Raman} scattering in the kinetic regime},
    journal = {Physics of Plasmas},
    volume = {32},
    number = {9},
    pages = {092102},
    year = {2025},
    month = {09},
    issn = {1070-664X},
    doi = {10.1063/5.0277771},
    fullurl = {https://doi.org/10.1063/5.0277771},
    fulleprint = {https://pubs.aip.org/aip/pop/article-pdf/doi/10.1063/5.0277771/20683023/092102_1_5.0277771.pdf},
}

@article{Winjum18,
  title = {Mitigation of stimulated {Raman} scattering in the kinetic regime by external magnetic fields},
  author = {Winjum, B. J. and Tsung, F. S. and Mori, W. B.},
  journal = {Phys. Rev. E},
  volume = {98},
  issue = {4},
  pages = {043208},
  numpages = {5},
  year = {2018},
  month = {Oct},
  publisher = {American Physical Society},
  doi = {10.1103/PhysRevE.98.043208},
  fullurl = {https://link.aps.org/doi/10.1103/PhysRevE.98.043208}
}

@article{Montgomery15,
    author = {Montgomery, D. S. and Albright, B. J. and Barnak, D. H. and Chang, P. Y. and Davies, J. R. and Fiksel, G. and Froula, D. H. and Kline, J. L. and MacDonald, M. J. and Sefkow, A. B. and Yin, L. and Betti, R.},
    title = {Use of external magnetic fields in hohlraum plasmas to improve laser-coupling},
    journal = {Physics of Plasmas},
    volume = {22},
    number = {1},
    pages = {010703},
    year = {2015},
    month = {01},
    issn = {1070-664X},
    doi = {10.1063/1.4906055},
    fullurl = {https://doi.org/10.1063/1.4906055},
    fulleprint = {https://pubs.aip.org/aip/pop/article-pdf/doi/10.1063/1.4906055/13673706/010703_1_online.pdf},
}

@article{Froula07,
  title = {Quenching of the Nonlocal Electron Heat Transport by Large External Magnetic Fields in a Laser-Produced Plasma Measured with Imaging {Thomson} Scattering},
  author = {Froula, D. H. and Ross, J. S. and Pollock, B. B. and Davis, P. and James, A. N. and Divol, L. and Edwards, M. J. and Offenberger, A. A. and Price, D. and Town, R. P. J. and Tynan, G. R. and Glenzer, S. H.},
  journal = {Phys. Rev. Lett.},
  volume = {98},
  issue = {13},
  pages = {135001},
  numpages = {4},
  year = {2007},
  month = {Mar},
  publisher = {American Physical Society},
  doi = {10.1103/PhysRevLett.98.135001},
  fullurl = {https://link.aps.org/doi/10.1103/PhysRevLett.98.135001}
}

@article{shi2018laser,
  title={Laser-plasma interactions in magnetized environment},
  author={Shi, Yuan and Qin, Hong and Fisch, Nathaniel J},
  journal={Physics of Plasmas},
  volume={25},
  number={5},
  year={2018},
  publisher={AIP Publishing}
}

@book{michel2023introduction,
  title={Introduction to laser-plasma interactions},
  author={Michel, Pierre},
  year={2023},
  publisher={Springer Nature}
}

@article{Bailly-Grandvaux2024,
  title = {Impact of strong magnetization in cylindrical plasma implosions with applied {B-field} measured via {x-ray} emission spectroscopy},
  author = {Bailly-Grandvaux, M. and Florido, R. and Walsh, C. A. and P{\'e}rez-Callejo, G. and Beg, F. N. and Bradford, P. and Gigosos, M. A. and Mancini, R. C. and McGuffey, C. and Suzuki-Vidal, F. and Vlachos, C. and Santos, J. J.},
  journal = {Phys. Rev. Res.},
  volume = {6},
  issue = {1},
  pages = {L012018},
  numpages = {8},
  year = {2024},
  month = {Jan},
  publisher = {American Physical Society},
  doi = {10.1103/PhysRevResearch.6.L012018},
  fullurl = {https://link.aps.org/doi/10.1103/PhysRevResearch.6.L012018}
}

@article{Sio21,
    author = {Sio, H. and Moody, J. D. and Ho, D. D. and Pollock, B. B. and Walsh, C. A. and Lahmann, B. and Strozzi, D. J. and Kemp, G. E. and Hsing, W. W. and Crilly, A. and Chittenden, J. P. and Appelbe, B.},
    title = {Diagnosing plasma magnetization in inertial confinement fusion implosions using secondary deuterium-tritium reactions},
    journal = {Review of Scientific Instruments},
    volume = {92},
    number = {4},
    pages = {043543},
    year = {2021},
    month = {04},
    issn = {0034-6748},
    doi = {10.1063/5.0043381},
    fullurl = {https://doi.org/10.1063/5.0043381},
    fulleprint = {https://pubs.aip.org/aip/rsi/article-pdf/doi/10.1063/5.0043381/14032636/043543_1_online.pdf},
}

@article{Knauer10,
    author = {Knauer, J. P. and Gotchev, O. V. and Chang, P. Y. and Meyerhofer, D. D. and Polomarov, O. and Betti, R. and Frenje, J. A. and Li, C. K. and Manuel, M. J.-E. and Petrasso, R. D. and Rygg, J. R. and Séguin, F. H.},
    title = {Compressing magnetic fields with high-energy lasers},
    journal = {Physics of Plasmas},
    volume = {17},
    number = {5},
    pages = {056318},
    year = {2010},
    month = {05},
    issn = {1070-664X},
    doi = {10.1063/1.3416557},
    fullurl = {https://doi.org/10.1063/1.3416557},
    fulleprint = {https://pubs.aip.org/aip/pop/article-pdf/doi/10.1063/1.3416557/14922820/056318_1_online.pdf},
}

@article{Hohenberger12,
    author = {Hohenberger, M. and Chang, P.-Y. and Fiksel, G. and Knauer, J. P. and Betti, R. and Marshall, F. J. and Meyerhofer, D. D. and Séguin, F. H. and Petrasso, R. D.},
    title = {Inertial confinement fusion implosions with imposed magnetic field compression using the {OMEGA} Laser},
    journal = {Physics of Plasmas},
    volume = {19},
    number = {5},
    pages = {056306},
    year = {2012},
    month = {03},
    issn = {1070-664X},
    doi = {10.1063/1.3696032},
    fullurl = {https://doi.org/10.1063/1.3696032},
    fulleprint = {https://pubs.aip.org/aip/pop/article-pdf/doi/10.1063/1.3696032/13991202/056306_1_online.pdf},
}

@article{Chang11,
  title = {Fusion Yield Enhancement in Magnetized Laser-Driven Implosions},
  author = {Chang, P. Y. and Fiksel, G. and Hohenberger, M. and Knauer, J. P. and Betti, R. and Marshall, F. J. and Meyerhofer, D. D. and S{\'e}guin, F. H. and Petrasso, R. D.},
  journal = {Phys. Rev. Lett.},
  volume = {107},
  issue = {3},
  pages = {035006},
  numpages = {4},
  year = {2011},
  month = {Jul},
  publisher = {American Physical Society},
  doi = {10.1103/PhysRevLett.107.035006},
  fullurl = {https://link.aps.org/doi/10.1103/PhysRevLett.107.035006}
}

@article{Li2019,
title = {High-directional laser-plasma-induced {X-ray} source assisted by collimated electron beams in targets with a self-generated magnetic field},
journal = {Fusion Engineering and Design},
volume = {144},
pages = {193-201},
year = {2019},
issn = {0920-3796},
doi = {https://doi.org/10.1016/j.fusengdes.2019.05.001},
fullurl = {https://www.sciencedirect.com/science/article/pii/S0920379619306386},
author = {Huan Li and Xiaobin Tang and Shuang Hang and Yunpeng Liu and Junxu Mu and Wei Zhou},
}

@article{Filippov2021,
  author  = {Filippov, Evgeny D. and Makarov, Sergey S. and Burdonov, Konstantin F. and Yao, Weipeng and Revet, Guilhem and B{\'e}ard, Jerome and Bola{\~n}os, Simon and Chen, Sophia N. and Guediche, Amira and Hare, Jack and Romanovsky, Denis and Skobelev, Igor Yu. and Starodubtsev, Mikhail and Ciardi, Andrea and Pikuz, Sergey A. and Fuchs, Julien},
  title   = {Enhanced {X-ray} emission arising from laser-plasma confinement by a strong transverse magnetic field},
  journal = {Scientific Reports},
  year    = {2021},
  volume  = {11},
  number  = {1},
  pages   = {8180},
  doi     = {10.1038/s41598-021-87651-8},
  fullurl    = {https://doi.org/10.1038/s41598-021-87651-8},
  issn    = {2045-2322}
}

@article{Lezhnin26,
  title = {Expansion-Driven Self-Magnetization of High-Energy-Density Plasmas},
  author = {Lezhnin, K. V. and Totorica, S. R. and Griff-McMahon, J. and Medvedev, M. V. and Landsberger, H. and Diallo, A. and Fox, W.},
  journal = {Phys. Rev. Lett.},
  volume = {136},
  issue = {11},
  pages = {115101},
  numpages = {7},
  year = {2026},
  month = {Mar},
  publisher = {American Physical Society},
  doi = {10.1103/stmq-c433},
  fullurl = {https://link.aps.org/doi/10.1103/stmq-c433}
}

@article{Walsh_2021,
doi = {10.1088/1741-4326/ac25c1},
fullurl = {https://doi.org/10.1088/1741-4326/ac25c1},
year = {2021},
month = {sep},
publisher = {IOP Publishing},
volume = {61},
number = {11},
pages = {116025},
author = {Walsh, C.A. and Sadler, J.D. and Davies, J.R.},
title = {Updated magnetized transport coefficients: impact on laser-plasmas with self-generated or applied magnetic fields},
journal = {Nuclear Fusion},
}

@article{Farmer17,
    author = {Farmer, W. A. and Koning, J. M. and Strozzi, D. J. and Hinkel, D. E. and Berzak Hopkins, L. F. and Jones, O. S. and Rosen, M. D.},
    title = {Simulation of self-generated magnetic fields in an inertial fusion hohlraum environment},
    journal = {Physics of Plasmas},
    volume = {24},
    number = {5},
    pages = {052703},
    year = {2017},
    month = {05},
    issn = {1070-664X},
    doi = {10.1063/1.4983140},
    fullurl = {https://doi.org/10.1063/1.4983140},
    fullprint = {https://pubs.aip.org/aip/pop/article-pdf/doi/10.1063/1.4983140/15990488/052703_1_online.pdf},
}

@article{moody2022increased,
  author  = {Moody, J. D. and Pollock, B. B. and Sio, H. and Strozzi, D. J. and Ho, D. D.-M. and Walsh, C. A. and Kemp, G. E. and Lahmann, B. and Kucheyev, S. O. and Kozioziemski, B. and others},
  title   = {Increased Ion Temperature and Neutron Yield Observed in Magnetized Indirectly Driven $D_2$-Filled Capsule Implosions on the {National Ignition Facility}},
  journal = {Physical Review Letters},
  volume  = {129},
  number  = {19},
  pages   = {195002},
  year    = {2022},
  doi     = {10.1103/PhysRevLett.129.195002}
}

@article{schmit2014understanding,
  author  = {Schmit, P. F. and Knapp, P. F. and Gomez, M. R. and Slutz, S. A. and Aefkow, A. B. and Sinars, D. B. and Fitzpatrick, K. J. and Campbell, E. M.},
  title   = {Understanding Fuel Magnetization and Secondary Nuclear Reactions in Magnetized Liner Inertial Fusion},
  journal = {Physical Review Letters},
  volume  = {113},
  number  = {15},
  pages   = {155004},
  year    = {2014},
  doi     = {10.1103/PhysRevLett.113.155004}
}

@article{appelbe2021magnetic,
  author  = {Appelbe, B. and Velikovich, A. L. and Sherlock, M. and Walsh, C. and Crilly, A. and O'Neill, S. and Chittenden, J.},
  title   = {Magnetic field transport in propagating thermonuclear burn},
  journal = {Physics of Plasmas},
  volume  = {28},
  number  = {3},
  pages   = {032705},
  year    = {2021},
  doi     = {10.1063/5.0040161}
}

@article{oneill2025burn,
  author  = {O'Neill, S. T. and Appelbe, B. D. and Crilly, A. J. and Walsh, C. A. and Strozzi, D. J. and Moody, J. D. and Chittenden, J. P.},
  title   = {Burn propagation in magnetized high-yield inertial fusion},
  journal = {Physics of Plasmas},
  volume  = {32},
  number  = {2},
  pages   = {022703},
  year    = {2025},
  doi     = {10.1063/5.0242215}
}

@article{stamper1971spontaneous,
  author  = {Stamper, J. A. and Papadopoulos, K. and Sudan, R. N. and Dean, S. O. and McLean, E. A. and Dawson, J. M.},
  title   = {Spontaneous Magnetic Fields in Laser-Produced Plasmas},
  journal = {Physical Review Letters},
  volume  = {26},
  number  = {17},
  pages   = {1012--1015},
  year    = {1971},
  doi     = {10.1103/PhysRevLett.26.1012}
}

@book{stix1992waves,
  author    = {Stix, Thomas H.},
  title     = {Waves in Plasmas},
  publisher = {American Institute of Physics},
  address   = {New York},
  year      = {1992},
  isbn      = {978-0883188590}
}

\end{document}